# Decentralized Unlabeled Multi-Agent Navigation in Continuous Space[*]

Stepan Dergachev[1,2] [0000−0001−8858−2831] and Konstantin Yakovlev[2,1] [0000−0002−4377−321X]

[1] HSE University, Moscow, Russia
[2] Federal Research Center for Computer Science and Control of Russian Academy of Sciences, Moscow, Russia
{dergachev,yakovlev}@isa.ru

**Abstract.** In this work, we study the problem where a group of mobile agents needs to reach a set of goal locations, but it does not matter which agent reaches a specific goal. Unlike most of the existing works on this topic that typically assume the existence of the centralized planner (or controller) and limit the agents' moves to a predefined graph of locations and transitions between them, in this work we focus on the decentralized scenarios, when each agent acts individually relying only on local observations/communications and is free to move in arbitrary direction at any time. Our iterative approach involves agents individually selecting goals, exchanging them, planning paths, and at each time step choose actions that balance between progressing along the paths and avoiding collisions. The proposed method is shown to be complete under specific assumptions on how agents progress towards their current goals, and our empirical evaluation demonstrates its superiority over a baseline decentralized navigation approach in success rate (i.e. is able to solve more problem instances under a given time limit) and a comparison with the centralized TSWAP algorithm reveals its efficiency in minimizing trajectory lengths for mission accomplishment.

## 1 Introduction

Multi-agent navigation is a vital and non-trivial problem which arises in various domains such as swarm robotics, transportation systems, video-games etc. Generally, the problem asks to find a set of non-colliding trajectories (paths) for a group of agents operating in a shared workspace. Numerous modifications, setups and approaches for this problem exist. One of the most well-studied setups is when each agent is asked to

---

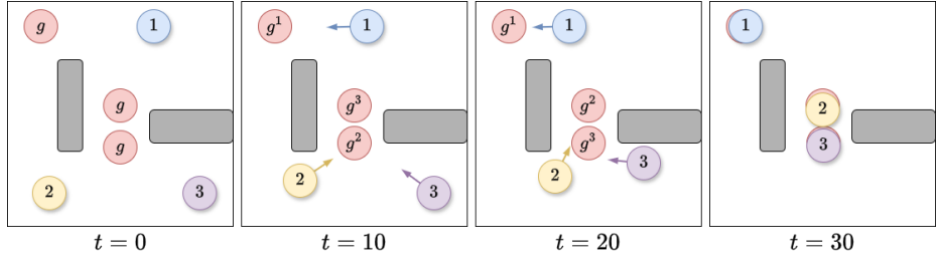

**Fig. 1.** An example of the unlabeled multi-agent navigation problem: the agents are denoted by blue, yellow, and violet circles (with numbers showing their identifiers), and the goals are represented by the red ones. Initially, at $t = 0$ (as shown in the upper-left pane), no fixed assignment of the goals to the agents is provided. The upper-right pane shows a possible intermediate state at $t = 10$, where each goal is already assigned to a particular agent (goal assignment is denoted via the superscript of the goal label), and each agent is moving towards its goal. At $t = 20$, the yellow and violet agents communicate and decide to exchange their goals ($g^3, g^2 \rightarrow g^2, g^3$) to efficiently solve the problem. At time moment $t = 30$, all agents reach their goals, and the problem is considered to be solved.

reach a specific goal location, i.e. the assignment of goals to agents is given as the problem input.

Another variant of this problem is known as unlabeled multi-agent navigation [14]. In this scenario, the agents are considered interchangeable, meaning that each agent is not strictly assigned to a particular goal. This is the specific problem that we investigate in this work.

Traditionally, methods for addressing (unlabeled) multi-agent navigation problems rely on a central controller that creates a common non-conflicting plan executed by all agents [16, 10, 20, 18, 15].

Interestingly, only a limited number of works delve into decentralized scenarios with restricted communication [12, 11, 9, 7]. However, even in these cases, there is an underlying assumption of initial consistent goal assignments for the agents, which may not hold in a fully decentralized setting. Furthermore, these methods do not account for the presence of static obstacles in the environment.

Table 1 provides a high-level comparison of the current methodologies capable of addressing the discussed problem. It is evident that the least explored area currently is decentralized navigation, especially when considering the absence of an initial consistent assignment, within a complex continuous workspace.

To mitigate these issues, we explore the decentralized unlabeled multi-agent navigation problem in a continuous workspace and propose an approach tailored to solve it. In summary, our approach combines individual pathfinding, an iterative selection of actions that enable each agent to move safely along the path, and a goal-exchanging procedure (see Fig. 1). The latter is particularly crucial as it significantly enhances the efficiency of the navigation process. We operate under the assumption that each agent lacks access to global information and chooses its actions based on local observations and local communication with other agents. This makes the problem of achieving the consistent (and effective) goal assignment more challenging as compared to the centralized case.

**Table 1.** A comparative table for existing key unlabeled navigation methods. The following designations are used: **Dec.** – a method is decentralized; **Cont.** – a method utilizes continuous workspace; **Obst.** – a method is able to work in complex environment; **Inconsist.** – a method works in the unavailability of an initially consistent goals assignment

|      | Dec. | Cont. | Obst. | Inconsist. |
|------|------|-------|-------|------------|
| [10] | -    | -     | +     | NA         |
| [15] | -    | +     | +     | NA         |
| [12] | +    | +     | -     | -          |
| [9]  | +    | +     | -     | +          |
| Ours | +    | +     | +     | +          |

We prove that the introduced method offers guarantees of completeness under well-defined conditions and restrictions (regarding how agents advance towards their respective goals). Our experiments indicate that the proposed approach can effectively address a wide range of challenging problem instances. Additionally, we contrast it with the centralized algorithm, TSWAP [10], which relies on the discretization of the workspace. Our method is shown to produce significantly shorter resultant trajectories, with an average reduction of 65% (and even more for specific configurations).

## 2   Related Works

The following topics are especially relevant to our work: *multi-agent path finding (MAPF), anonymous/unlabeled MAPF, unlabeled motion planning, trajectory generation and goal assignment* etc.

The problem of *Multi-Agent Path Finding (MAPF)* is actively being investigated, and various formulations of the MAPF problem are considered in the works devoted to this topic. A significant portion of studies on MAPF assume the presence of a centralized controller and utilize a discretized workspace representation, such as a grid-graph or roadmap. Numerous techniques exist to address MAPF, with some aiming for provably optimal solutions [13] or bounded- suboptimal outcomes [2]. However, these methods often struggle to scale effectively when dealing with a large number of agents. Conversely, for scenarios, where finding a solution quickly takes precedence over cost efficiency, rule-based solvers can be employed [5]. Prioritized planning might offer a potential middle ground between solution cost and performance, often being fast and scalable while yielding near-optimal solutions [3]. Nevertheless, prioritized planning is generally not guaranteed to find a complete solution.

One of the variations of MAPF is the *Anonymous* or *Unlabeled* MAPF, where the goals are not initially assigned to the agents [16, 10, 20]. Unlike classical MAPF, AMAPF always has a solution [10]. Most of the works devoted to this problem make assumptions about the central controller and the discrete workspace representation. Meanwhile, there are centralized approaches [17, 1, 15] operating in a continuous space. However, they either do not take into account the static obstacles [17] or impose significant restrictions on inputs, such as the minimum relative location of the starting and goal positions from the static obstacles [1] and/or the minimum relative location of the starting and goal positions to the static obstacles [15].

On the other hand, there are only a limited number of decentralized approaches to the unlabeled motion planning problem. An adaptation of the centralized solver CAPT to a decentralized scenario is presented in [17]. Unfortunately, it does not guarantee the absence of collisions between agents. The method described in [12] (based on [11]) relies on switching between two policies: one for goal assignment and the other for collision avoidance. Although this method provides guarantees in obstacle-free environments, it is not suitable for environments with obstacles. Additionally, the goal assignment strategy of this method requires considering $N!$ (where $N$ is the number of agents) different assignments, which can become infeasible for large $N$.

Another rapidly growing research direction in this area involves the use of neural networks and multi-agent reinforcement learning for both centralized and decentralized navigation and goal assignment, as seen in [9, 8, 6, 7]. However, these methods require extensive training and often struggle with generalizing to problem instances that even slightly differ from those used during training. Moreover, they typically lack strong theoretical guarantees.

## 3   Problem Statement

Consider a 2-dimensional workspace, $W \subseteq \mathbb{R}^2$, consisting of the free space, $W_{free} \subseteq W$ and the obstacles, $W_{obs} = W \setminus W_{free}$. A set of $n$ homogeneous agents, $\mathcal{N}$, operates in the workspace, and a set $G$ of $n$ goals in $W_{free}$ is distinguished. We assume each agent $i \in \mathcal{N}$ is identified by a unique number (e.g., serial number of the robot). To simplify the further narrative, we will assume that the agent and the identifier are identical.

Let $T = 0, 1, 2, \ldots$, be the discrete time. At each time step, each agent chooses an action that brings it to the next location. We consider the following transition model for an individual agent: $\boldsymbol{p}_{t+1} = \boldsymbol{p}_t + \boldsymbol{u}_t$.

Here $\boldsymbol{p}_t = (x, y) \in W_{free}$ denotes the location of the agent at time $t$; $\boldsymbol{u}_t \in R^2$ is the chosen action, which is interpreted as the velocity of the agent, i.e. $\boldsymbol{u} = (dx, dy)$. The velocity is bounded: $\|\boldsymbol{u}\| \leq u_{max}$.

The considered transition model assumes that the agent is able to move into an arbitrary direction, as well as instantaneously accelerate or decelerate. We also assume that each action is executed perfectly.

A spatio-temporal path (trajectory) for an agent is a mapping: $\pi : T \to W_{free}$, that can be written as $\pi = \{\boldsymbol{p}_0, \boldsymbol{p}_1, \boldsymbol{p}_2, \ldots\}$. Alternatively, a trajectory can be represented as the sequence of the applied actions. In this work we are interested in converging trajectories, i.e. the ones where an agent reaches the neighborhood of a particular location and never moves away from it. Next, we will use the terms path and trajectory interchangeably.

A path for an individual agent is feasible *iff* it does not collide with a static obstacle and the velocity constraints are met. The paths for two distinct agents are said to be conflict-free if the agents following them never collide.

The time step, $t_{fin} \in T$, by which the agent reaches its final destination defines the duration of the path: $duration(\pi) = t_{fin}$. The total distance traveled by the agent during the execution of the path defines the length of the path $length(\pi) = \sum_{1}^{t_{fin}} ||\pi(t) - \pi(t-1)||$.

**The problem** now is to find a sequence of actions (a path) for each agent s.t. *(i)* each individual path is feasible and starts at the predefined start location and ends in (the neighborhood of) one of the predefined goals $g \in G$; *(ii)* all goal locations are reached; *(iii)* each pair of paths is conflict-free.

## 4    Unlabeled Multi-agent Navigation via Goal Exchanging

Our method can be seen as an extension of the solver for the labeled multi-agent navigation. Thus, we, first, briefly describe how the decentralized methods for the labeled case are typically organized and then switch to the unlabeled scenario.

### 4.1    Decentralized Multi-agent Navigation: Labeled Case

Typically, most of the decentralized methods for the labeled multi-agent navigation rely on two main ingredients: pathfinding (long-term planning) and collision avoidance (short-term decision-making). The first component provides a global reference path that circumnavigates the static obstacles, while the second is tailored to follow the reference path, e.g. by trying to reach the path's next way-point, and to avoid the local collisions (with both the static obstacles and the other agents) at the same time.

In this case, at the beginning of a mission, an agent finds a path to the predefined goal using information only about a static environment. After that, the agent starts moving along the constructed path, trying not to collide with other agents and static obstacles. In case the agent deviates significantly from the original reference path, a re-planning procedure might be triggered.

The exact algorithms for pathfinding and collision avoidance are domain specific. In what follows next, we rely on the assumption that such methods are available for us as the high-level routines `constructPath` and `computeAction`. The latter is assumed to produce an action that makes an agent progress towards the next waypoint on the path in a safe way.

```
Algorithm 1: DEC-UNAV
   Input: i – agent; W – workspace; p – position; G – set of all goals; u_max – action limit.
 1 g ← selectGoal(G, W, p); π ← constructPath(p, g, W);
 2 if path π not found then
 3  |  return fail;
 4 status ← moveToGoal; GR ← {};
 5 GR[g'] ← false : ∀g' ∈ G;
 6 while true do
 7  |  N ← identifyLocalAvailableAgents(); P, U ← receivePositionsControls(N);
 8  |  GA ← recieveGoalAssignment(N); AS ← recieveStatuses(N);
 9  |  GR[g'] ← ⋁_{j∈N} GR^j[g'] : ∀g' ∈ G;
10  |  g', status', GR ← goalUpdate(i, N, G, P, GA, AS, GR);
11  |  if goalUpdate failed then
12  |   |  return fail;
13  |  status ← status';
14  |  if (g ≠ g') or (p deviated from old path) then
15  |   |  π ← constructPath(p, g', W); g ← g'
16  |  u ← computeAction(P, U, W, π, u_max);
17  |  p ← applyAction(u);
```

**Fig. 2.** Pseudocode of decentralized unlabeled navigation algorithm.

### 4.2  From Labeled Scenario to the Unlabeled One

The navigation scheme described earlier can successfully cope with the labeled scenarios, where it is assumed that the unique goal is assigned to each of the agents. In this work, however, we do not assume this. Particularly, we assume that each agent is, indeed, capable of choosing a goal, but different agents may choose the same goal initially. Thus, if they do not change their decision, the instance will not be solved. To this end, we add a component that allows agents to change their decisions on which goals to pursue based solely on local inter-agent communication. In what follows next we, first, put this procedure into the context of the (individual) navigation algorithm and then describe the goal update procedure in details.

### 4.3  Navigation Pipeline (Individual Agent)

In this section, we would like to give a general description of the suggested **DEC**entralized **U**nlabeled **NAV**igation scheme (*DEC-UNAV*). Note that we basically rely on the same components (namely, path planning, which is de- noted as constructPath procedure, and collision avoidance, which is denoted as computeAction procedure) and ideas that are used in labeled case (described in section 4.1). However, there are a number of key differences that allow us to find a solution in the unlabeled scenario.

The main contribution is that the current agent assigns a goal to itself, after which the goal can be changed during execution. For the initial assignment, either a random selection (but path existence check should be performed) or the closest location (in terms of path length) to the agent can be used. Goal updating produced by `goalUpdate` procedure. This procedure is necessary to eliminate the inconsistency of global goal assignment and to increase the efficiency of solving the problem. The detailed description of the suggested mechanism is provided further in section 4.4.

```
Algorithm 2: goalUpdate procedure
Input: i – agent; N – neighboring agents in priority order; G – set of all goals; P – agents'
       positions; GA – goals assignment; AS – agents' statuses; GR – reached goals table.
Output: g' – new goal of i; status' – new status of i; GR – updated reached goals table.
1  for j ∈ N do
       /* Reassign goal if agent j's goal is reached by another agent          */
2      if (AS[j] = moveToGoal) and (GR[GA[j]] = true) then
3          GA[j] ← g : closest to P[j] in G and GR[g] = false;
4          if pathLen(P[j], GA[j])= ∞ then
5              return fail
       /* Update agent j's status upon reaching its goal                        */
6      if (AS[j] ≠ reached) and (near(P[j], GA[j])) then
7          AS[j] ← reached; GR[GA[j]] ← true;
8      for k ∈ N and j > k do
           /* Resolve goal assignment inconsistencies between agents j and k    */
9          if GA[j] = GA[k] then
10             g' ← g : closest to P[j] in G and GR[g] = false;
11             if (AS[j] ≠ reached) and (GA[j] ≠ g') then
12                 GA[j] ← g' ;
13             else
14                 GA[k] ← g : closest to P[k] in G\GA[j] and GR[g] = false;
15                 if pathLen(P[k], GA[k])= ∞ then
16                     return fail
17                 AS[k] ← moveToGoal;
18         else
               /* The assignment is consistent, but goals swap may be beneficial */
19             if (AS[j], AS[k] ≠ reached) and (GR[GA[k]] = false) then
20                 if swappable(P[j], GA[j], P[k], GA[k]) then
21                     swap(GA[j], GA[k]) ;
22     G ← G\GA[j];
23 return GA[i], AS[i], GR ;
```

**Fig. 3.** Pseudocode of goal updating procedure.

Agents should store and send information about their goals through local communication, as well as store the information about the goals of neighboring agents at every time step using a special table/dictionary of goals $GA$. Let denote the goal of the neighboring agent $j$ stored in memory of agent at current time step as $GA[j]$.

Next, each agent must store and exchange information whether it has reached (or not) some goal. Let's indicate status as $reached$ for agents who have achieved goals. Otherwise, the agent's status is denoted as $moveToGoal$. Similar to goals, the agent also must receive information about the statuses of neighboring agents and store it using a special table/dictionary $AS$. The status of neighboring agent $j$ we denote as $AS[j]$.

Finally, agents store (and exchange) information about whether a particular goal has been reached by some of the agents using a table $GR$, which persists throughout execution. Initially, each goal $g' \in G$ is marked as $GR[g'] = false$, indicating none have been achieved. Upon receiving confirmation that a goal $g'$ has been reached, agents update $GR[g']$ to $true$.

The information stored in tables $GA, AS$ and $GR$ is necessary for elimination of inconsistency, and the procedure **goalUpdate** may update values in these tables.

It is important to note that the goal exchange procedure is not initiated every iteration of the algorithm; specific conditions triggering its execution are detailed in Section 4.6.

For a detailed implementation, refer to the pseudocode provided in Algorithm 1 (see Fig. 2).

### 4.4 Goal Updating

The most important component that allows you to correctly perform the task is the goal update mechanism. Recall that the operations indicated below are performed only on the basis of information that is known to a certain group of agents sharing information and available for communication with each other. The key ideas for suggested technic are presented below.

The first suggestion is for each agent $j \in N$ to verify whether its current goal $g$ has already been achieved by another agent. If, after exchanging information, it is discovered that the goal has indeed been achieved by the other agent, the agent $j$ should choose an alternative goal. The second suggestion aims to restore consistent goal assignment within the connected group $N$. To achieve this, we propose prioritizing agents using unique identifiers (e.g., serial numbers of the robots) and ensuring that for each agent $j \in N$, no lower-priority agent $k$ is moving towards the same goal. The third suggestion involves exchanging goals between agents if doing so can enhance the effectiveness of the solution and simplify collision avoidance.

Algorithm 2 (see Fig. 3) iterates over each agent $j$ in the current connected group $N$ to handle goal reassignment and status updates. If it is discovered that agent $j$'s assigned goal has been reached by another agent (line 2), the procedure recalculates $j$'s goal to the closest available element in $G$ that has not yet been reached (line 3), ensuring the path to the new goal is feasible. If no such goal exists, the procedure terminates with a failure (lines 4-5).

After that, we check if an agent $j$ gets close enough to its goal, its status is updated to *reached* and the corresponding entry in $GR$ is set to *true*, indicating the goal has been achieved (lines 6-7).

Next, inconsistencies in goal assignment between agent $j$ and lower-priority agents $k$ are detected and resolved by reassigning goals based on proximity to available goals in $G$ (lines 8-21). Two cases should be considered: *(i)* if the agent $j$ has not reached its goal $GA[j]$ and there is a vacant goal $g'$ that is closer to $j$, then $j$'s goal should be updated; *(ii)* otherwise, the assignment for a lower-priority agent k should be changed. Separately, it is worth considering the scenario where both agents $j$ and $k$ believe they have achieved the same goal $g$. This situation may occur if the first agent to reach the goal was significantly displaced during collision avoidance and then returned to $g$. In such cases, the goal $g$ will be reassigned to the agent with higher priority, $j$, while the lower-priority agent $k$ must choose a new goal.

Additionally, the algorithm facilitates goal swapping between agents j and k when it is beneficial for solution quality or collision avoidance, contingent upon both agents not having reached their respective goals (lines 19-21).

We introduce a criterion named **swappable** to check the possibility of exchanging different goals between two agents to ensure the theoretical properties presented in section 4.6. The exchange will occur only if *(i)* the higher priority agent reduces its distance to the goal and *(ii)* the total distance for both agents is also reduced. More formally, agents $j$ and $k$, where $j > k$, with positions $p$ and $p'$, should exchange their goals $g$ and $g'$ only if the following conditions are met: $pathLen(p, g) > pathLen(p, g')$ and $pathLen(p, g) + pathLen(p', g') > pathLen(p, g') + pathLen(p', g)$.

After processing each agent $j$, goals that have been assigned are removed from $G$, ensuring agents choose from remaining unassigned goals in subsequent iterations (line 22).

Based on the procedure, updated goal assignments, agent statuses, and reached goals table are obtained. Using these updated data, a new goal and status for the agent $i$, who executed the procedure, are computed (line 23).

It is important to highlight that in practical implementations using existing collision avoidance methods, deadlock situations can arise, particularly when one agent occupies a goal directly in the path of another agent. To address this challenge, we propose an additional mechanism for exchanging goals between two adjusted agents. This exchange occurs when one agent is in close proximity to its goal and the path of another agent goes through with it.

### 4.5 Centralized Case

Although our work is mainly focused on decentralized scenario, for completeness in the study of the topic, we also consider a centralized scenario. In centralized case, it is assumed that there exists a controller that obtains full information about all agents and is able to communicate with them flawlessly to issue the commands, which action to take. In this setting, our task is to decide the actions to be taken by the agents at each time step, based on the full knowledge of the environment, current goals assignment and agents' states.

The algorithm for **C**entralized **U**nlabeled **NAV**igation (**C-UNAV**) is built on the same principles that underlie the decentralized version, but includes several differences in goal exchanging and navigation pipeline.

First of all, it is assumed that the global consistent goal assignment is defined. So, there is no need to bring it into a consistent form, it is only necessary to exchange goals. Secondly, the goal exchanging procedure may process all agents in the scenario (instead of separate subgroups). We suggest considering each pair of agents to determine the need for swapping. And the primary determinant for the swap of goals between two agents is the minimization of the distance that these agents must overcome to reach their goals. More formally, agents $i, j$ with positions $p$ and $p'$ should exchange their goals $g$ and $g'$ only if $pathLen(p, g) + pathLen(p', g') > pathLen(p, g') + pathLen(p', g)$.

### 4.6 Theoretical Analysis

In this section, we will outline several theoretical properties of the proposed algorithms. For the sake of brevity, we will provide sketches of proofs that explain the main ideas without exhaustive details.

Firstly, we introduce the following definitions and remarks. Let us denote the global goals assignment for agents of $\mathcal{N}$ as $\overline{GA} \subset G \times \mathcal{N}$, where $\forall i \in \mathcal{N} \; \exists (i, g) \in \overline{GA}$. The global goals assignment $\overline{GA}$ is termed *consistent* if each goal $g \in G$ is assigned to exactly one agent $i \in \mathcal{N}$. Next, let $S$ denote the total length of the paths agents traverse from their current locations to their assigned goal locations.

*Remark 1.* We assume that the `goalUpdate` procedure (Alg. 2, see Fig. 3) is periodically executed for each agent $i \in \mathcal{N}$. Between executions, we assume no collisions occur with other agents or static obstacles. Additionally, we expect the value of $pathLen(p^i, g^i)$ to decrease by at least some $\varepsilon \in \mathbb{R}^+$, except in cases where the agent reaches proximity to the goal.

**Theorem 1**. *If a solution of the problem exists, then in the process of navigation using the Alg. 1(see Fig. 2) there is a time moment $t$ when the global goal assignment $\overline{GA}$ becomes consistent.*

*Proof.* Observe that the number of reached goals in $GR$ table of any agent does not decrease. Suppose that there is no such time moment after which the goal assignment remains consistent. But there must be such a moment of time t' after which the number of reached goals in $GR$ does not change for each agent in $\mathcal{N}$. Consider two cases:

1. **Case 1:** *There is at least one agent, that did not reach a goal (i.e. status of agent is $moveToGoal$).*
   Let $i$ be the agent with the highest priority among agents with $moveToGoal$ status. Let $p$ be the position and $g$ be the goal of agent $i$. The condition at line 2 of Alg. 2 cannot be met (otherwise, it would imply that the $GR$ table for agent $i$ has changed relative to the previous time step, leading to a contradiction). Therefore, agent $i$'s goal can change only after completing lines 9 or 19 of Alg. 2. Consequently, the value $pathLen(p, g)$ cannot increase during the execution of Alg. 2. According to Remark 1, the agent $i$ will either reach its goal and change $GR$, or while approaching the goal $g$, the agent $i$ will encounter another agent who has already achieved it, thereby changing $GR$. This leads to a contradiction.
2. **Case 2:** *All agents believe that they have reached goals, but at least two agents $i, j$ have the same goal $g$.*
   In this scenario, neither agent will change its goal until they meet each other at ***g*** (according to Remark 1). Once agents $i$ and $j$ meet, the agent with lower priority will change its goal, allowing us to apply the statement of this theorem again until the assignment becomes consistent, or until some agent's table changes. Hence, this is a contradiction.

**Theorem 2.** *All goals will be achieved by the agents using Algorithm 1 (see Fig. 2), if possible, or one or more agents will terminate task execution (with respect to Remark 1).*

*Proof.* Firstly, consider the case of solution existence. According to Theorem 1, there exists a moment when the goal assignment becomes consistent. We aim to prove that if the assignment becomes consistent, it will remain so.

Consider running Alg. 2 (see Fig. 3) for a connected subgroup of agents. The conditions at lines 2 and 9 will never be true because with a consistent assignment, two agents cannot have the same goal. Therefore, changes in agents' goals can only occur during goal swaps, which maintain assignment consistency.

Furthermore, after each run of the swap operation inside the `goalUpdate` procedure for a subgroup of agents with a consistent assignment, the total sum of path lengths $S$ either decreases or remains unchanged. According to Remark 1, S decreases by at least some ε between executions of `goalUpdate`. Since $S \geq 0$, there will be a moment when all agents are sufficiently close to their goals, ensuring all goals are considered achieved.

Now, suppose no solution exists. This implies there exists a non-empty subset of agents for whom paths exist only to a subset of goals, where the number of such goals is fewer than the number of agents in the subset. Consequently: *(i)* either the condition at line 2 in Alg. 1 is true (indicating at least one agent cannot reach any goal), or *(ii)* during goal updating, an agent in the subset fails to find an achievable goal not selected by higher-priority agents (lines 4-5 and 15-16 in Alg. 2).

*Remark 2*. Note that similar reasoning as in Theorem 2 is also applicable to the centralized scenario (*C-UNAV* algorithm), under the conditions that *(i)* the initial assignment of goals is consistent and feasible, and *(ii)* the goal exchange procedure is executed periodically. Between these executions, no collisions occur, and the value of $S$ decreases by at least some $δ \in \mathbb{R}^+$, except when all agents are already as close as possible to their goals.

## 5  Experimental Evaluation

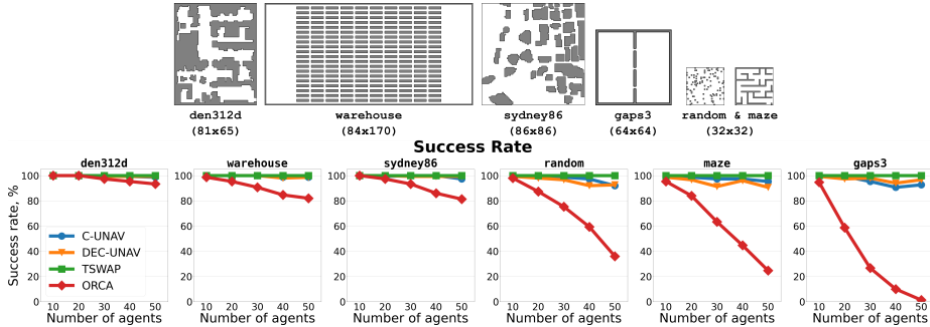

**Fig. 4.** Maps on which were used in experiments (top row) and *success rates* of the algorithms on corresponding maps. The higher is the better (bottom row).

We implemented the proposed method in C++[2] and evaluated its efficiency across various scenarios. Our implementation uses *Theta\** [4] for pathfinding and the *ORCA* [19] algorithm for generating safe actions in continuous spaces.

---

[2] https://github.com/PathPlanning/Decentralized-Unlabeled-Navigation

We utilized six distinct grid maps in our experiments (see top row of Fig. 4). The first map, *den312d*, measures 81 × 65 and includes several large rooms and wide corridors. The second map, *warehouse*, is characterized by a single large room with regularly spaced elongated obstacles, spanning 84 × 170 in size. The third map, *sydney86*, represents an outdoor city environment and measures 86 × 86. The fourth map, *random*, is a 32 × 32 grid with randomly placed obstacles occupying 10% of its cells. The fifth map, *maze*, is also 32 × 32 and features a maze-like structure. The last map, $gaps3$, measures 64 × 64 and consists of two open areas separated by a wall with three narrow passages. Maps *den312d*, *warehouse*, *random*, and *maze* were sourced from the *MovingAI* MAPF benchmark [16].

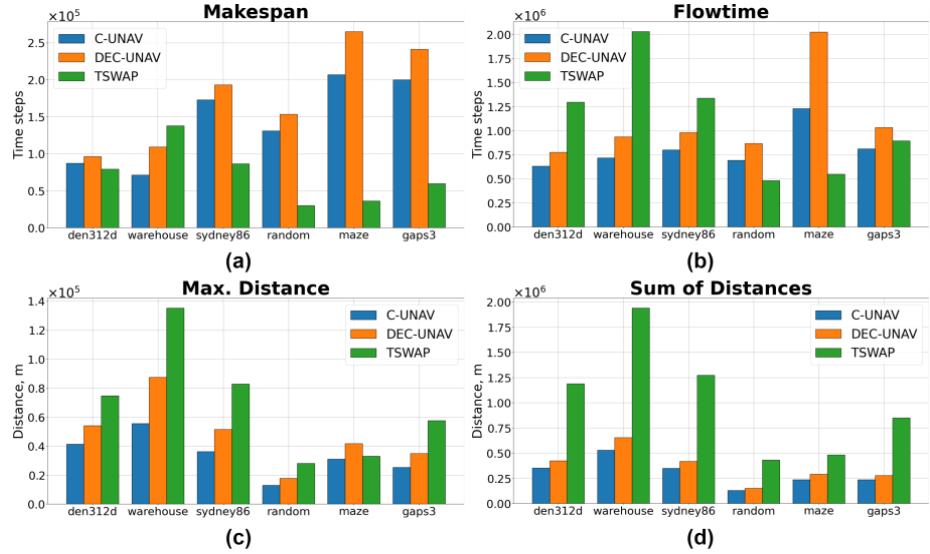

**Fig. 5.** The sum of (a) *makespan*, (b) *flowtime*, (c) *maxdist* and (d) *sumdist* values for evaluated algorithms on different maps. The result on a specific instance was added to the sum only if all the algorithms successfully solved it. The lower is the better.

The experimental scenarios were generated as follows: for each map, we created 150 instances, with each instance containing 50 start/goal locations for the agents. Each start/goal pair was positioned at the center of a random cell on the grid. The number of agents, $n$, varied from 10 to 50 in increments of 10, with the first $n$ start/goal pairs used in each scenario. Each agent has the maximum velocity of 0.1 cells per time step and modeled as a disk with a radius of 0.3 of the cell width. For collision avoidance purposes, a safety margin was added, using a disk radius of 0.49.

We compared our proposed method, *DEC-UNAV*, against a baseline approach (*ORCA*), which combines *ORCA* for collision avoidance and *Theta*$^*$ for path planning but does not incorporate goal exchanging. Initially, agents controlled by *DEC-UNAV* independently selected the closest goals. The goal exchanging procedure was triggered every 20 time steps.

Additionally, we evaluated a centralized version of our method, *C-UNAV*, and the centralized discrete AMAPF algorithm, *TSWAP* [10], using the same grid maps. Agent sizes were chosen to be less than $\sqrt{2}/4$, enabling simultaneous movement through neighboring cells without collisions. For consistency, the same initial goal assignment for *C-UNAV, ORCA* and *TSWAP* algorithms was employed.

Each simulation run was limited to a maximum of 20,000 time steps, except for the TSWAP algorithm, where the limit was 2,000 steps. This adjustment accounts for the discrete nature of the TSWAP algorithm, with each step equivalent to 10 steps of algorithms operating in continuous space. A run was deemed a *failure* if, by the maximum time step limit: *(i)* at least one agent failed to reach its goal position; *(ii)* at least one collision occurred between agents or with static obstacles; *(iii)* a deadlock situation was detected, defined as the average speed of all agents over the last 1,000 time steps being below 0.0001 cells per time step. Conversely, a run was considered a *success* if all goals were occupied by agents within the specified time limit.

We tracked the following indicators: **success rate** – the percentage of instances that result with *success* out of all tasks; **makespan** – the maximum duration among all agents' trajectories, measured as the time step when all goals were occupied by agents; **flowtime** – the sum of trajectory durations for all agents, where an agent's trajectory duration is the time step when it reaches one of goals and remains there without further movement; **maxdist** – the longest trajectory length among all agents; **sumdist** – the sum of trajectory lengths for all agents. When computing *makespan* and *flowtime* for *TSWAP*, each step of this algorithm was set to 10 time steps (since the maximum speed of the agents was set to 0.1 cell per time step).

The success rates of *DEC-UNAV, C-UNAV, TSWAP,* and *ORCA* are depicted at the bottom of Fig. 4 (the higher – the better). Overall, the centralized discrete approach, *TSWAP*, demonstrated the highest success rates, solving all tasks across all maps. However, the proposed method, particularly the decentralized variant *DEC-UNAV*, achieved comparable success rates to the centralized counterpart. Notably, the proposed method, incorporating goal exchange, consistently outperformed the baseline (*ORCA*) on all maps. This difference was especially pronounced on maps featuring narrow passages (*maze*, *random*, and *gaps3*), where the baseline struggled due to multiple agents converging in confined spaces. In contrast, the proposed method effectively mitigated collisions through goal exchanging. Nonetheless, even with centralized goal exchanging, achieving a 100% success rate on these maps remained elusive. This challenge likely stems from the inherent difficulty of coordinating multiple agents through narrow passages using the *ORCA* collision avoidance algorithm, where agents act independently, potentially impeding each other's movements.

Next, we proceeded to evaluate the solution quality of *DEC-UNAV, C-UNAV*, and *TSWAP*. Fig. 5 illustrates the aggregated values of quality indicators across various maps.

Fig. 5-a demonstrates that *TSWAP* typically achieves solutions with shorter durations, despite being constrained to cardinal moves on discrete cells. However, in terms of total duration across all solutions (Fig. 5-b), the proposed methods generally outperform *TSWAP*. Specifically, on maps with ample open space and devoid of narrow passages, the proposed method consistently generates solutions of superior quality

compared to *TSWAP*. For example, on the *den312d* map, the flowtime of *DEC-UNAV* is approximately 65% lower than that of *TSWAP*, and on the warehouse map, *DEC-UNAV* achieves more than double the reduction in flowtime compared to *TSWAP*. Notably, the differences in both flowtime and makespan between centralized and decentralized algorithms are negligible, except for the maze map where the decentralized approach exhibits around 60% higher flowtime, likely due to the map's complex structure necessitating more frequent navigation through difficult areas.

The difference between the proposed method and *TSWAP* can also be attributed to the collision avoidance method often choosing velocities well below the maximum limit, particularly near obstacles and narrow passages. Consequently, although most agents reach their goals promptly, some may encounter bottlenecks that significantly increase makespan without similarly affecting flowtime. In contrast, the discrete nature of *TSWAP* ensures agents consistently move at maximum speed, albeit potentially inefficiently near obstacles and passages.

This hypothesis is further supported by metrics related to distance traveled (Fig. 5-c,d). Total trajectory lengths are consistently greater for *TSWAP*, with differences ranging from 65% on the maze map to 179%-205% on other maps. Similar trends are observed when considering the maximum distance traveled, although the differences are less pronounced, with the decentralized approach performing slightly worse than *TSWAP* on the maze map. Meanwhile, the centralized *C-UNAV* achieves modest reductions (15%-23% for *sumdist* and 30%-57% for *maxdist*) in both maximum and total distances traveled compared to *DEC-UNAV*.

In conclusion, our experiments demonstrate that the proposed approach effectively addresses challenging instances and solves more tasks compared to the baseline. Moreover, the solutions obtained are either superior or slightly inferior to those achieved by a centralized discrete algorithm in terms of solution duration and trajectory lengths. This underscores the practical efficacy of the suggested goal exchanging techniques and their contribution to the efficiency of the proposed solvers.

## 6    Conclusion

This paper addresses the problem of multi-agent navigation with interchangeable agents and introduces *DEC-UNAV*, a decentralized method tailored for solving this challenge. Central to our approach is a goal exchanging procedure that operates while agents navigate towards their goals. We have demonstrated that under certain assumptions governing agent movement towards goals, our approach achieves completeness. Additionally, we extended our method to a centralized version, *C-UNAV*.

We validated these methods through a series of simulated experiments across various maps, comparing them against: *(i)* a baseline method without goal exchange, and *(ii)* the state-of-the-art centralized discrete AMAPF algorithm, *TSWAP*. Our results indicate that our methods successfully solve challenging tasks where the baseline fails. Moreover, the solutions obtained generally match or exceed those achieved by *TSWAP*, demonstrating improvements in solution quality in terms of trajectories lengths by 65% to 205%.

Future work will focus on relaxing assumptions about agent movement towards goals. Another direction involves adapting our method to scenarios where agents are grouped into teams, each with distinct goals and interchangeable agents. Incorporating kinodynamic constraints and conducting experiments with real robots are also promising avenues for further research.

**Acknowledgments.** The reported study was supported by the Ministry of Science and Higher Education of the Russian Federation under Project 075-15-2024-544.

**Disclosure of Interests.** The authors have no competing interests to declare that are relevant to the content of this article.